\documentclass[12pt]{amsart}

\allowdisplaybreaks

\usepackage{amssymb}
\usepackage[mathscr]{eucal}

\numberwithin{equation}{section}

\newtheorem{theorem}{Theorem}[section]
\newtheorem{proposition}[theorem]{Proposition}
\newtheorem{corollary}[theorem]{Corollary}
\newtheorem{lemma}[theorem]{Lemma}

\newcommand{\C}{\mathbb{C}}
\newcommand{\p}{\partial}
\newcommand{\half}{\tfrac{1}{2}}
\newcommand{\CP}{\mathbb{CP}}
\newcommand{\<}{\langle}
\renewcommand{\>}{\rangle}

\newcommand{\UU}{\mathcal{U}}
\newcommand{\VV}{\mathcal{V}}
\newcommand{\CL}{\mathcal{L}}
\renewcommand{\AA}{\mathcal{A}}

\newcommand{\Tau}{\mathcal{O}}
\DeclareMathOperator{\Tr}{Tr}
\newcommand{\MM}{\mathcal{C}}
\newcommand{\RR}{\mathcal{R}}
\renewcommand{\eth}{D}
\newcommand{\Dil}{\mathcal{D}}
\newcommand{\G}{\psi}

\newcommand{\Mbar}{\overline{\mathcal{M}}}
\newcommand{\PHI}{\mathcal{F}}
\newcommand{\Det}{\Delta}
\renewcommand{\tt}{\tilde{t}}
\newcommand{\Euler}{\mathcal{E}}
\newcommand{\BB}{\mathsf{B}}

\begin{document}

\title[Topological gravity in genus $2$]
{Topological gravity in genus $2$ \\ with two primary fields}

\author{Tohru Eguchi, Ezra Getzler and Chuan-Sheng Xiong}

\address{Department of Physics, Faculty of Science, University of Tokyo,
Tokyo 113, Japan}

\email{eguchi@phys.s.u-tokyo.ac.jp}

\address{RIMS, Kyoto University, Kitashirakawa Oiwake-cho, Sakyo-ku, Kyoto
602, Japan}

\address{Department of Mathematics, Northwestern University, Evanston, IL
60208, USA}

\email{getzler@math.northwestern.edu}

\address{Department of Physics, Beijing University, Beijing 100871, China}

\email{xiong@ibm320h.phy.pku.edu.cn}

\begin{abstract}
We calculate the genus $2$ correlation functions of two-dimensional
topological gravity, in a background with two primary fields, using the
genus $2$ topological recursion relations.
\end{abstract}

\maketitle

In this paper, we calculate the genus $2$ correlation functions of
two-dimensional topological gravity in a background with two primary fields
$\Tau_0$ and $\Tau_1$; this extends the work of Eguchi, Yamada and Yang
\cite{EYY}, who considered the case of the $A_2$-model.

The most interesting example of such a theory is the Gromov-Witten theory
of $\CP^1$; in this case, there is a rigorous construction of the
correlation functions (see Manin \cite{Manin}). For $\CP^1$, our
calculation may be made into a rigorous proof. One of our motivations was
to confirm that the resulting potential is consistent with the Toda
conjecture of Eguchi and Yang \cite{EY}.

In the general case, in order to complete the proof, we must use the
equation $L_1Z=0$, which is part of the Virasoro conjecture of Eguchi, Hori
and Xiong \cite{EHX}. We verify that the Virasoro conjecture then holds in
genus $2$ for these models.

Our results agree with those of Dubrovin and Zhang \cite{DZ}, who use the
method of Eguchi and Xiong \cite{EX}; in particular, they use the Virasoro
constraints $L_nZ=0$, $n\le10$.

\section{Topological recursion relations}

\subsection{Notation}
The correlators of the theory are denoted $\< \tau_{k_1,a_1} \dots
\tau_{k_n,a_n} \>_g$. We denote $\tau_{0,a}$ by $\Tau_a$. The labels on the
primaries are fixed in such a way that the puncture operator is
$\Tau_0$. Let $\eta_{ab}$ be the intersection form, $\eta^{ab}$ its
inverse, and let $\Tau^a=\eta^{ab}\Tau_b$. In the case of two primaries,
the intersection form equals $\eta_{ab}=\delta_{a+b,1}$.

Let $\PHI_g$ be the genus $g$ potential on the large phase space:
\begin{equation} \label{Phi}
\PHI_g = \sum_{n=0}^\infty \frac{1}{n!} \sum_{\substack{ k_1\dots
k_n \\ a_1 \dots a_n }} t^{a_1}_{k_1} \dots t^{a_n}_{k_n} \< \tau_{k_1,a_1}
\dots \tau_{k_n,a_n} \>_g .
\end{equation}
we use the summation convention with respect to the indices $a_i$ labelling
the primaries.

Denote $\p/\p t^a_k$ by $\p_{k,a}$. The vector field $\p=\p_{0,0}$,
corresponding to the puncture operator $\Tau_0$, plays a special role in
the theory. The partial derivatives of the potential $\PHI_g$ are denoted
$$
\<\< \tau_{k_1,a_1} \dots \tau_{k_n,a_n} \>\>_g = \p_{k_1,a_1} \dots
\p_{k_n,a_n} \PHI_g .
$$

\newpage

\subsection{The topological recursion relation in genus $0$}
The simplest example of a topological recursion relation is obtained by
taking the relation $\psi_1=0$ on the zero-dimensional moduli space
$\Mbar_{0,3}$. The resulting topological recursion relation is the equation
\begin{equation} \label{trr0}
\<\<\tau_{k,a}\tau_{\ell,b}\tau_{m,c}\>\>_0 = \<\<\tau_{k-1,a}\Tau^d\>\>_0
\<\<\Tau_d\tau_{\ell,b}\tau_{m,c}\>\>_0 .
\end{equation}

Let $\Theta$ be the power series
$$
\Theta(z)_a^b = \delta_a^b + \sum_{k=0}^\infty z^{k+1}
\<\<\tau_{k,a}\Tau^b\>\>_0 ;
$$
it is an orthogonal matrix, in the sense that
$\Theta^{-1}(z)=\Theta^*(-z)$. Let $\UU$ be the matrix with components
$\UU_a^b = \<\<\Tau_a\Tau^b\>\>_0$. The topological recursion relation
\eqref{trr0} with $m=0$ may be rewritten as
\begin{equation} \label{TRR0}
\p_{k,a}\Theta(z) = z\,\Theta(z)\,\p_{k,a}\UU .
\end{equation}
Let $\p_a(z) = \sum_{k=0}^\infty z^k \, \p_{k,a}$, and define vector fields
$\{\eth_{k,a}\mid k\ge 0 \}$ on the large phase space by
$$
\eth_a(z) = \sum_{k=0}^\infty z^k \eth_{k,a} = \Theta^{-1}(z)_a^b \,
\p_b(z) .
$$
For example, $\eth_{0,a}=\p_{0,a}$ and
$\eth_{1,a}=\p_{1,a}-\UU_a^b\p_{0,b}$.
\begin{lemma} \label{dU}
We have $\eth_a(z)\UU=\p_{0,a}\UU$ and
$\eth_a(z)\Theta(w)=w\Theta(w)\p_{0,a}\UU$.
In particular, $\eth_{k,a}\UU=\eth_{k,a}\Theta=0$ if $k>0$.
\end{lemma}
\begin{proof}
It follows easily from \eqref{trr0} that $\eth_a(z)\UU=\p_{0,a}\UU$; since
$$
\eth_a(z)\Theta(w)=w\Theta(w)\eth_a(z)\UU,
$$
the result follows.
\end{proof}

\begin{corollary}
The vector fields $\eth_{k,a}$ and $\eth_{\ell,b}$ commute if both $k$
and $\ell$ are positive, while
$$
[\eth_{k,a},\p_{0,b}] = \<\<\Tau_a\Tau_b\Tau^c\>\>_0 \, \eth_{k-1,c} .
$$
\end{corollary}
\begin{proof}
By Lemma \ref{dU},
\begin{align*}
\eth_a(w) \eth_b(z) &= \eth_a(w) \, \Theta^{-1}(z)_b^c \, \p_c(z) \\
&= \Theta^{-1}(z)_b^c \, \eth_a(w) \, \p_c(z) -
z\,\<\<\Tau_a\Tau_b\Tau^c\>\>_0 \, \eth_c(z) \\
&= \Theta^{-1}(z)_b^c \, \Theta^{-1}(w)_a^d \, \p_d(w) \, \p_c(z) -
z\,\<\<\Tau_a\Tau_b\Tau^c\>\>_0 \, \eth_c(z) .
\end{align*}
It follows that $[\eth_a(w),\eth_b(z)]=\<\<\Tau_a\Tau_b\Tau^c\>\>_0 (
w\,\eth_c(w) - z\,\eth_c(z) )$.
\end{proof}

This corollary leads to an algorithm for the calculation of
$\eth_a(z)\<\<\Tau_{a_1}\dots\Tau_{a_n}\>\>_g$ by induction on $n$ in terms
of $\eth_a(z)\PHI_g$, using the formula
\begin{equation} \label{delta}
\eth_a(z)\<\<\Tau_{a_1}\dots\Tau_{a_n}\>\>_g = \sum_{i=1}^n \p_{0,a_1}\dots
[\eth_a(z),\p_{0,a_i}] \dots \p_{0,a_n} \PHI_g + \p_{0,a_1}\dots\p_{0,a_n}
\eth_a(z)\PHI_g .
\end{equation}

\subsection{The string equation in genus $0$ and coordinates on the large
phase space}

The genus $0$ string equation says that $\CL_{-1}\PHI_0+\half \eta_{ab}
t^a_0 t^b_0=0$, where $\CL_{-1}$ is the vector field
$$
\CL_{-1} = \sum_{k=0}^\infty t_{k+1}^a \p_{k,a} - \p_{0,0} .
$$
The string equation implies the following lemma.
\begin{lemma} \label{small}
The restriction of $\p\UU$ to the small phase space $\{t^a_k=0 \mid k>0\}$
equals the identity, while for $n>1$, the restriction of $\p^n\UU$ to the
small phase space vanishes.
\end{lemma}
\begin{proof}
The vector fields $\p_{0,a}$ commute with $\CL_{-1}$; it follows that
$$
\CL_{-1}\UU_{ab} = \CL_{-1}\p_{0,a}\p_{0,b}\PHI_0 =
\p_{0,a}\p_{0,b}\CL_{-1}\PHI_0 = - \eta_{ab} .
$$
Written out explicitly, this equation says that
$$
\p\UU_a^b = \delta_a^b + \sum_{k=0}^\infty t^c_{k+1}
\<\<\Tau_a\Tau^b\tau_{k,c}\>\>_0 .
$$
Applying the operator $\p^{n-1}$, $n>0$, we obtain
$$
\p^n\UU_a^b = \sum_{k=0}^\infty t^c_{k+1}
\p^{n-1}\<\<\Tau_a\Tau^b\tau_{k,c}\>\>_0 .
$$
The lemma is an immediate consequence of these formulas.
\end{proof}

In conjunction with the genus $0$ topological recursion relation, this
implies the following theorem.
\begin{theorem} \label{main}
Let $u^a=\p\<\<\Tau^a\>\>_0$. The functions $u_n^a=\p^n u^a$, $n\ge0$, form
a coordinate system in a neighbourhood of the small phase space, and
\begin{equation} \label{Du}
D_a(z) = \sum_{n=0}^\infty ((\p+z\,\p\UU)^n\p\UU)_a^b \, \frac{\p}{\p
u^b_n} .
\end{equation}
\end{theorem}
\begin{proof}
Since $u^b=\UU^b_0$, Lemma \ref{dU} implies that
$$
\eth_a(z) u^b_n = \bigl( \Theta^{-1}(z) \, \p^n \, \Theta(z) \, \p\UU
\bigr)_a^b = \bigl( ( \Theta^{-1}(z) \cdot \p \cdot \Theta(z) )^n \, \p\UU
\bigr)_a^b .
$$
Since $\Theta^{-1}(z)\cdot\p\cdot\Theta(z)=\p+z\,\p\UU$ by \eqref{TRR0}, we
conclude that $\eth_a(z)u^b_n=((\p+z\,\p\UU)^n\p\UU)_a^b$.

By Lemma \ref{small}, the restriction of $(\p+z\,\p\UU)^n\p\UU$ to the
small phase space equals $z^n$. It follows that the restriction of
$\eth_{k,a}u^b_n$ to the small phase space equals $\delta_{k,n}\delta_a^b$;
hence the functions $u^a_n$ form a coordinate system in a neighbourhood of
the small phase space.
\end{proof}

Note that $(\p+z\,\p\UU)^n\p\UU=z^{-1}p_{n+1}(z\p\UU)$, where
$p_{n+1}(f)=(\p+f)^nf$ is the $(n+1)$st Fa\`a~di~Bruno polynomial.

\begin{corollary} \label{vanish}
If $\eth_{k,a}f=0$ for $k>n$, then $\p f/\p u^a_k=0$ for $k>n$.
\end{corollary}

\begin{corollary}
In terms of the coordinates $u^a_n$, the small phase space $\{t^a_k=0\mid
k>0\}$ is the submanifold
$$
u^a_n = \begin{cases} \delta^a_0 & n=1 , \\ 0 & n>1 . \end{cases}
$$
\end{corollary}

Theorem \ref{main} shows that the large phase space may be defined for any
Frobenius manifold $M$, as the infinite jet space $J^\infty M$ (i.e.\
Dubrovin's ``loop space''). This is seen by rewriting the matrix
$\p\UU_b^a$ as $\p u^c \, \AA_{bc}^a$, where
\begin{equation} \label{AA}
\AA_{bc}^a = \frac{\p\UU^a_b}{\p u^c}
\end{equation}
is the tensor describing the product on the tangent bundle of $M$.

An attractive feature of the vector fields $\eth_{k,a}$ is that they
commute with $\CL_{-1}$:
\begin{align*}
[\CL_{-1},\eth_a(z)] &= [\CL_{-1},\Theta^{-1}(z)_a^b\,\p_b(z)] =
[\CL_{-1},\Theta^{-1}(z)_a^b] \, \p_b(z) + \Theta^{-1}(z)_a^b \,
[\CL_{-1},\p_b(z)] \\
&= ( z\,\Theta^{-1}(z)_a^b ) \, \p_b(z) - \Theta^{-1}(z)_a^b \,
( z \, \p_b(z) ) = 0 .
\end{align*}

By the genus $0$ string equation, $\CL_{-1}u^a_n$ vanishes for $n>0$, while
$\CL_{-1}u^a=-\delta^a_0$: it follows that in the coordinate system
$\{u^a_n\}$, the vector field $\CL_{-1}$ is given by the formula
$$
\CL_{-1} = - \frac{\p}{\p u^0} .
$$
In the coordinate system $\{u^a_n\}$, the string equation
$\CL_{-1}\PHI_g=0$ says that $\PHI_g$ is independent of $u^0$.

Lemma \ref{small} shows that $\p\UU$ is invertible in a neighbourhood of
the small phase space: denote its inverse by $\MM$. We also see that its
determinant $\Delta=\det(\p\UU)$ equals $1$ on the small phase space.

\subsection{The topological recursion relation in genus $1$}

We now illustrate the way in which use of the vector fields $\eth_{k,a}$
simplifies the discussion of topological recursion relations, using as an
example the topological recursion relation in genus $1$:
\begin{equation} \label{trr1}
\<\<\tau_{k,a}\>\>_1 = \<\<\tau_{k-1,a}\Tau^b\>\>_0 \<\<\Tau_b\>\>_1 +
\tfrac{1}{24} \<\<\tau_{k-1,a}\Tau_b\Tau^b\>\>_0 .
\end{equation}
Multiplying by $z^k$ and summing over $k$, we obtain
$$
\p_a(z) \PHI_1 = \Theta(z)_a^b \<\<\Tau_b\>\>_1 + \tfrac{1}{24} \, z \,
\p_a(z) \Tr(\UU) ,
$$
hence, by Lemma \ref{dU},
$$
\eth_a(z) \PHI_1 = \<\<\Tau_b\>\>_1 + \tfrac{1}{24} \, z \, \eth_a(z)
\Tr(\UU) = \<\<\Tau_b\>\>_1 + \tfrac{1}{24} \, z \, \p_{0,a} \Tr(\UU) .
$$
This may be written as the sequence of differential equations
\begin{equation} \label{TRR1}
\eth_{k,a}\PHI_1 = \begin{cases} \tfrac{1}{24} \, \p_{0,a} \Tr(\UU) & k=1 ,
\\ 0 & k>1 .
\end{cases}
\end{equation}

The equations \eqref{TRR1} have the particular solution
$\frac{1}{24}\log(\Det)$. Let $\G = \PHI_1 - \tfrac{1}{24} \log(\Det)$; we
see that $D_{k,a}\G=0$ for all $k>0$. Hence, by Corollary \ref{vanish},
$\G$ depends only on the coordinates $u^a$; by the string equation, it is
independent of $u^0$. In this way, we recover a result of Dijkgraaf and
Witten~\cite{DW}: there is a function $\G$ of the coordinates $\{u^a\}$
such that $\PHI_1=\tfrac{1}{24}\log(\Det)+\G$.

\subsection{The dilaton equation}
The dilaton equation is another important constraint on the potentials of
topological gravity. Let $\Dil$ be the vector field
$$
\Dil = \p_{1,0} - \sum_{k=0}^\infty t_k^a \p_{k,a} .
$$
The dilaton equation says that
$$
\Dil\PHI_g = \begin{cases}
(2g-2)\PHI_g , & g\ne1 , \\
\chi/24 , & g=1 , \\
\end{cases}$$
where $\chi$ is the Euler characteristic of the background.
\begin{proposition}
In the coordinate system $\{u^a_n\}$, the dilaton vector field $\Dil$ equals
$$
\Dil = \sum_{n=1}^\infty n \, u^a_n \, \frac{\p}{\p u^a_n} .
$$
\end{proposition}
\begin{proof}
By the genus $0$ dilaton equation $\Dil\PHI_0=-2\PHI_0$, we have $\Dil
u^a_n = n \, u^a_n$, and the formula for $\Dil$ follows.
\end{proof}

\section{The $A_2$ and $\CP^1$ models in genus $2$}

In genus $2$, there are two topological recursion relations
\cite{genus2}. The first is
\begin{equation} \label{trr21}
\begin{aligned}
\<\<\tau_{k,a}\>\>_2 &= \<\<\tau_{k-1,a}\Tau^b\>\>_0 \<\<\Tau_b\>\>_2
+ \<\<\tau_{k-2,a}\Tau^b\>\>_0 \bigl( \<\<\tau_{1,b}\>\>_2 -
\<\<\Tau_b\Tau^c\>\>_0 \<\<\Tau_c\>\>_2 \bigr) \\
&+ \<\<\tau_{k-2,a}\Tau^b\Tau^c\>\>_0 \bigl( \tfrac{7}{10} \, \<\<\Tau_b\>\>_1
\<\<\Tau_c\>\>_1 + \tfrac{1}{10} \, \<\<\Tau_b\Tau_c\>\>_1 \bigr) \\
&+ \tfrac{13}{240} \<\<\tau_{k-2,a}\Tau^b\Tau^c\Tau_c\>\>_0 \<\<\Tau_b\>\>_1
- \tfrac{1}{240} \<\<\tau_{k-2,a}\Tau^b\>\>_1 \,
\<\<\Tau_b\Tau^c\Tau_c\>\>_0 \\
&+ \tfrac{1}{960} \<\<\tau_{k-2,a}\Tau^b\Tau_b\Tau^c\Tau_c\>\>_0 .
\end{aligned}
\end{equation}
Using the topological recursion relations in genus $0$ and $1$,
\eqref{trr21} may be rewritten as the sequence of differential equations
\begin{equation} \label{TRR21}
\eth_{k,a}\PHI_2 = \RR_{k,a} ,
\end{equation}
where
$$
\RR_{k,a} = \begin{cases}
\<\<\Tau_a\Tau_b\Tau_c\>\>_0 \bigl( \tfrac{7}{10}
\, \<\<\Tau^b\>\>_1 \<\<\Tau^c\>\>_1 + \tfrac{1}{10} \,
\<\<\Tau^b\Tau^c\>\>_1 \bigr) & \\
+ \tfrac{13}{240} \<\<\Tau_a\Tau_b\Tau_c\Tau^c\>\>_0 \<\<\Tau^b\>\>_1
- \tfrac{1}{240} \<\<\Tau_a\Tau^b\>\>_1 \<\<\Tau_b\Tau_c\Tau^c\>\>_0 & k=2
, \\
+ \tfrac{1}{960} \<\<\Tau_a\Tau_b\Tau^b\Tau_c\Tau^c\>\>_0 & \\[5pt]
\<\<\Tau_a\Tau_b\Tau_c\>\>_0 \bigl( \tfrac{1}{20} 
\<\<\Tau^b\>\>_1 \<\<\Tau^c\Tau^d\Tau_d\>\>_0 +
\tfrac{1}{480} \<\<\Tau^b\Tau^c\Tau^d\Tau_d\>\>_0 \bigr) & k=3 , \\
+ \tfrac{1}{1152} \<\<\Tau_a\Tau^b\Tau^c\Tau_c\>\>_0
\<\<\Tau_b\Tau^d\Tau_d\>\>_0 & \\[5pt]
\tfrac{1}{1152} \<\<\Tau_a\Tau^b\Tau^c\>\>_0
\<\<\Tau_b\Tau_c\Tau^d\>\>_0 \<\<\Tau_d\Tau^e\Tau_e\>\>_0 & k=4 , \\[5pt]
0 & k>4 .
\end{cases}$$

The other topological recursion relation in genus $2$ is,
\begin{multline} \label{trr22}
\<\<\tau_{k,a}\tau_{\ell,b}\>\>_2 = \<\<\tau_{k,a}\Tau_c\>\>_2
\<\<\Tau^c\tau_{\ell-1,b}\>\>_0 + \<\<\tau_{k-1,a}\Tau_c\>\>_0
\<\<\Tau^c\tau_{\ell,b}\>\>_2 \\
\begin{aligned}
{} &- \<\<\tau_{k-1,a}\Tau_c\>\>_0 \<\<\tau_{\ell-1,b}\Tau_d\>\>_0
\<\<\Tau^c\Tau^d\>\>_2 \\ &+ 3 \<\<\tau_{k-1,a}\tau_{\ell-1,b}\Tau^c\>\>_0
\bigl( \<\<\tau_{1,c}\>\>_2 - \<\<\Tau_c\Tau^d\>\>_0 \<\<\Tau_d\>\>_2
\bigr) \\
& + \tfrac{13}{10} \<\<\tau_{k-1,a}\tau_{\ell-1,b}\Tau_c\Tau_d\>\>_0
\<\<\Tau^c\>\>_1 \<\<\Tau^d\>\>_1 \\
{}& + \tfrac{4}{5} \bigl( \<\<\tau_{k-1,a}\Tau_c\>\>_1 \<\<\Tau_d\>\>_1 +
\tfrac{1}{24} \<\<\tau_{k-1,a}\Tau_c\Tau_d\>\>_1 \bigr)
\<\<\tau_{\ell-1,b}\Tau^c\Tau^d\>\>_0 \\
{}& + \tfrac{4}{5} \<\<\tau_{k-1,a}\Tau^c\Tau^d\>\>_0
\bigl( \<\<\tau_{\ell-1,b}\Tau_c\>\>_1 \<\<\Tau_d\>\>_1 +
\tfrac{1}{24} \<\<\tau_{\ell-1,b}\Tau_c\Tau_d\>\>_1 \bigr) \\
&- \tfrac{4}{5} \<\<\tau_{k-1,a}\tau_{\ell-1,b}\Tau_c\>\>_0 \bigr(
\<\<\Tau^c\Tau_d\>\>_1 \<\<\Tau^d\>\>_1 +
\tfrac{1}{24} \<\<\Tau^c\Tau_d\Tau^d\>\>_1 \bigr) \\
{}& + \tfrac{1}{48} \<\<\tau_{k-1,a}\Tau_c\Tau_d\Tau^d\>\>_0
\<\<\Tau^c\tau_{\ell-1,b}\>\>_1
+ \tfrac{1}{48} \<\<\tau_{k-1,a}\Tau_c\>\>_1
\<\<\tau_{\ell-1,b}\Tau^c\Tau_d\Tau^d\>\>_0 \\
& + \tfrac{23}{240}
\<\<\tau_{k-1,a}\tau_{\ell-1,b}\Tau_c\Tau_d\Tau^d\>\>_0 \<\<\Tau^c\>\>_1
- \tfrac{1}{80} \<\<\tau_{k-1,a}\tau_{\ell-1,b}\Tau_c\>\>_1
\<\<\Tau^c\Tau^d\Tau_d\>\>_0 \\ {}& + \tfrac{7}{30}
\<\<\tau_{k-1,a}\tau_{\ell-1,b}\Tau_c\Tau_d\>\>_0
\<\<\Tau^c\Tau^d\>\>_1 + \tfrac{1}{576}
\<\<\tau_{k-1,a}\tau_{\ell-1,b}\Tau_c\Tau^c\Tau_d\Tau^d\>\>_0 .
\end{aligned}
\end{multline}
Taking $k$ and $\ell$ equal to $1$ and using the topological recursion
relations in genus $0$ and $1$, we obtain the system of differential
equations
\begin{equation} \label{TRR22}
\eth_{1,1,a,b}\PHI_2 = \RR_{1,1,a,b} ,
\end{equation}
where $\eth_{1,1,a,b} = \eth_{1,a} \eth_{1,b} - 3 \,
\<\<\Tau_a\Tau_b\Tau^c\>\>_0 \, \eth_{1,c}$, and
\begin{align*}
\RR_{1,1,a,b} & = \tfrac{13}{10} \<\<\Tau_a\Tau_b\Tau_c\Tau_d\>\>_0
\<\<\Tau^c\>\>_1 \<\<\Tau^d\>\>_1 \\
{}& + \tfrac{4}{5} \bigl( \<\<\Tau_a\Tau_c\>\>_1 \<\<\Tau_d\>\>_1 +
\tfrac{1}{24} \<\<\Tau_a\Tau_c\Tau_d\>\>_1 \bigr)
\<\<\Tau_b\Tau^c\Tau^d\>\>_0 \\
{}& + \tfrac{4}{5} \<\<\Tau_a\Tau^c\Tau^d\>\>_0
\bigl( \<\<\Tau_b\Tau_c\>\>_1 \<\<\Tau_d\>\>_1 +
\tfrac{1}{24} \<\<\Tau_b\Tau_c\Tau_d\>\>_1 \bigr) \\
&- \tfrac{4}{5} \<\<\Tau_a\Tau_b\Tau_c\>\>_0 \bigr(
\<\<\Tau^c\Tau_d\>\>_1 \<\<\Tau^d\>\>_1 +
\tfrac{1}{24} \<\<\Tau^c\Tau_d\Tau^d\>\>_1 \bigr) \\
{}& + \tfrac{1}{48} \<\<\Tau_a\Tau_c\Tau_d\Tau^d\>\>_0
\<\<\Tau^c\Tau_b\>\>_1+ \tfrac{1}{48} \<\<\Tau_a\Tau_c\>\>_1
\<\<\Tau_b\Tau^c\Tau_d\Tau^d\>\>_0 \\
& + \tfrac{23}{240} \<\<\Tau_a\Tau_b\Tau_c\Tau_d\Tau^d\>\>_0
\<\<\Tau^c\>\>_1 - \tfrac{1}{80} \<\<\Tau_a\Tau_b\Tau_c\>\>_1
\<\<\Tau^c\Tau^d\Tau_d\>\>_0 \\
{}& + \tfrac{7}{30} \<\<\Tau_a\Tau_b\Tau_c\Tau_d\>\>_0
\<\<\Tau^c\Tau^d\>\>_1 + \tfrac{1}{576}
\<\<\Tau_a\Tau_b\Tau_c\Tau^c\Tau_d\Tau^d\>\>_0 .
\end{align*}

We now specialize to the case of the $A_2$ model. In this model, there are
two primary fields $\Tau_0$ and $\Tau_1$, with intersection form
$\eta_{ab}=\delta_{a+b,1}$. Denote the associated coordinates
$u=\<\<\Tau_0\Tau_0\>\>_0=\p^2\PHI_0$ and $v=\<\<\Tau_0\Tau_1\>\>_0$. The
matrix $\UU$ is given by the formula
$$
\UU = \begin{bmatrix} \UU_0^0 & \UU_0^1 \\ \UU_1^0 & \UU_1^1 \end{bmatrix}
= \begin{bmatrix} v & u \\ u^2 & v \end{bmatrix} ,
$$
and $\PHI_1=\tfrac{1}{24}\log(\Det)$. As was shown by Eguchi, Yamada and
Yang \cite{EYY}, the genus $2$ potential of the $A_2$-model is given by the
formula
\begin{equation} \label{A2}
\begin{aligned}
\PHI_2 &= \tfrac{1}{1152} \, \p^2\<\<\Tau_a\Tau_b\Tau_c\Tau_d\>\>_0 \,
\MM^{ab} \MM^{cd} \\
&- \tfrac{1}{1152} \, \p^2\<\<\Tau_a\Tau_b\>\>_0 \,
\p\<\<\Tau_c\Tau_d\Tau_e\Tau_f\>\>_0 \, \MM^{ac} \MM^{bd} \MM^{ef} \\
&- \tfrac{1}{360} \, \p^2\<\<\Tau_a\Tau_b\Tau_c\>\>_0 \,
\p\<\<\Tau_d\Tau_e\Tau_f\>\>_0 \, \MM^{ad} \MM^{be} \MM^{cf} \\
&+ \tfrac{1}{360} \, \p^2\<\<\Tau_a\Tau_b\>\>_0 \,
\p\<\<\Tau_c\Tau_d\Tau_e\>\>_0 \, \p\<\<\Tau_f\Tau_g\Tau_h\>\>_0 \,
\MM^{ac} \MM^{bf} \MM^{dg} \MM^{eh} .
\end{aligned}
\end{equation}
It may be checked that this function solves the equations \eqref{TRR21} and
\eqref{TRR22}.

For an arbitrary theory of topological gravity, let $\PHI_{2,0}$ be the
function on the large phase space given by formula \eqref{A2}. For all
theories of topological gravity for which we know the genus $2$ potential,
the function $\PHI_{2,0}$ appears to be a major contribution to this
potential.

We now turn to the case of $\CP^1$. As in the $A_2$-model, there are two
primary fields $\Tau_0$ and $\Tau_1$, with intersection form
$\eta_{ab}=\delta_{a+b,1}$. Again, denote the associated coordinates by
$u=\<\<\Tau_0\Tau_0\>\>_0=\p^2\PHI_0$ and $v=\<\<\Tau_0\Tau_1\>\>_0$. The
matrix $\UU$ is now given by the formula
$$
\UU = \begin{bmatrix} v & u \\ e^u & v \end{bmatrix} ,
$$
and $\PHI_1=\tfrac{1}{24}\log(\Det)-\tfrac{1}{24}u$. 

The correlators $\<\tau_{1,a_1}\Tau_{a_2}\dots\Tau_{a_n}\>_2$ and
$\<\Tau_{a_1}\Tau_{a_2}\dots\Tau_{a_n}\>_2$ vanish in the
$\CP^1$-model for dimensional reasons. It follows that the following
solution to the equations \eqref{TRR21} and \eqref{TRR22} is the genus
$2$ potential:
\begin{align} \label{CP1}
\PHI_2 = \PHI_{2,0} &- \tfrac{1}{480} \, \p^3\,\<\<\Tau_a\Tau_b\>\>_0 \,
\MM^{ab} + \tfrac{7}{5760} \, \p^3\,\<\<\Tau_a\>\>_0 \,
\p^2\<\<\Tau_b\>\>_0 \, \MM^{ab} \\ & + \tfrac{11}{5760} \,
\p^2\<\<\Tau_a\Tau_b\>\>_0 \, \p^2 \, \<\<\Tau_c\Tau_d\>\>_0 \, \MM^{ac} \,
\MM^{bd} . \notag
\end{align}
The three additional terms reflect the fact that, unlike in the
$A_2$-model, the function $\psi(u)=-\tfrac{1}{24}u$ is nonzero in the
$\CP^1$-model.

The Toda conjecture of Eguchi and Yang (\cite{EY}, \cite{EHY}, \cite{P})
provides conjectural formulas for the functions $\<\<\Tau_1\Tau_1\>\>_g$,
$g>0$, of the $\CP^1$-model:
$$
\sum_{g=0}^\infty \lambda^{2g} \<\<\Tau_1\Tau_1\>\>_g = \exp \biggl(
\frac{2}{\lambda^2} \bigl( \cosh(\lambda\p)-1 \bigr) \, \sum_{g=0}^\infty
\lambda^{2g} \PHI_g \biggr) .
$$
In genus $2$, this yields the equation
\begin{equation} \label{toda}
\<\<\Tau_1\Tau_1\>\>_2 = e^u \bigl( \p^2\PHI_2 + \tfrac{1}{12} \p^4\PHI_1 +
\tfrac{1}{360} \p^6\PHI_0 + \half ( \p^2\PHI_1 + \tfrac{1}{12} \p^4\PHI_0
)^2 \bigr) .
\end{equation}
It is easily checked, using the explicit formula formula for $\PHI_2$, that
this equation holds.

\section{Models with two primaries}

In this section, we consider topological gravity in a general background
with two primary fields $\Tau_0$ and $\Tau_1$, and intersection form
$\eta_{ab}=\delta_{a+b,1}$. It is not clear to what extent such a model,
even if it possesses a consistent loop expansion, corresponds to a physical
theory: it may be that only the $A_2$ and $\CP^1$-models are physical
theories. The fact that our equations remain consistent in this setting is
nevertheless very suggestive.

Denote the associated coordinates $u=\<\<\Tau_0\Tau_0\>\>_0$ and
$v=\<\<\Tau_0\Tau_1\>\>_0$. The genus~$0$ sector is characterized by the
function $\<\<\Tau_1\Tau_1\>\>_0$; by the string equation, this is a
function of $u$ alone, and we denote it by $\phi(u)$. The matrix $\UU$ is
given by the formula
$$
\UU = \begin{bmatrix} v & u \\ \phi(u) & v \end{bmatrix} .
$$

In this section, the correlation functions
$\<\<\tau_{k_1,a_1}\dots\tau_{k_n,a_n}\>\>_g$ are assumed to have the
following form: they are holomorphic functions of $\{(v,u)\in\C^2\mid
u\notin(-\infty,0]\}$, Laurent polynomials in $\Det$, and polynomial in the
remaining coordinates $\{\p^nv,\p^nu \mid n>0 \}$.

There is a universal differential equation \cite{elliptic} in topological
gravity relating the potentials $\PHI_0$ and $\PHI_1$. In the case of two
primary fields, this equation says that
\begin{equation} \label{elliptic}
\tfrac{1}{24} \phi'''+\phi'' \G'-2 \, \phi' \G'' = 0 .
\end{equation}
It turns out that this equation is also the necessary and sufficient
condition for the system of equations \eqref{TRR21} and \eqref{TRR22} to
have a solution. The necessity follows from the formula
$$
\eth_{1,1,0,0} \RR_{2,0} - \eth_{2,0} \RR_{1,1,0,0} = \tfrac{2}{15} \bigl(
\p u)^3 (4(\p v)^2+(\p u)^2\phi' \bigr) ( \tfrac{1}{24} \phi''' + \phi''
\G' - 2 \, \phi' \G'' ) .
$$

\begin{theorem}
Suppose that $\tfrac{1}{24} \phi''' + \phi'' \G' - 2 \, \phi' \G'' = 0$.
Then the equations \eqref{TRR21} and \eqref{TRR22} have the solution
$\PHI_{2,0}+\PHI_{2,1}$, where $\PHI_{2,0}$ is given by \eqref{A2}, and
\begin{align*}
\PHI_{2,1} &= \tfrac{1}{576} \bigl( ( \half \, \p\,\p_{0,a}\p_{0,b}\G
+ \tfrac45 \, \p\,\p_{0,a}\G \, \p_{0,b}\G ) \, \MM^{ab} \\
&+ \p^2\<\<\Tau_a\Tau_b\>\>_0 \, ( \tfrac65 \, \p_{0,c}\p_{0,d}\G 
- \tfrac{1}{10} \, \p_{0,c}\G \, \p_{0,d}\G ) \, \MM^{ac} \, \MM^{bd} \\
&+ ( \tfrac{7}{10} \, \p^2\<\<\Tau_a\Tau_b\Tau_c\>\>_0 \, \p_{0,d}\G
- \tfrac{3}{10} \, \p\<\<\Tau_a\Tau_b\Tau_c\>\>_0 \, \p\,\p_{0,d}\G ) \,
\MM^{ab} \, \MM^{cd} \\
&+ \p^2\<\<\Tau_a\Tau_b\>\>_0 \,
\p\<\<\Tau_c\Tau_d\Tau_e\>\>_0 \, \p_{0,f}\G \, ( \tfrac{3}{10} \,
\MM^{af} \, \MM^{bc} \, \MM^{de}
- \tfrac{23}{10} \, \MM^{ac} \, \MM^{bd} \, \MM^{ef} ) \\
&+ \tfrac{1}{10} \, (\p u)^4 \, \phi'' \, \G'' \, \Det^{-1} \bigr) .
\end{align*}
This solution may be characterized by the property that its restriction to
the small phase space, together with the restrictions of the functions
$\p_{1,a}(\PHI_{2,0}+\PHI_{2,1})$, vanish.
\end{theorem}

All of the terms in the formula for $\PHI_{2,0}+\PHI_{2,1}$ except the last
one $\frac{1}{5760}(\p u)^4 \, \phi'' \, \G'' \, \Det^{-1}$ are associated
to Feynman graphs with propagator $\MM$ and vertices
$\p^n\p_{a_1}\dots\p_{a_{k-2}}\UU_{a_{k-1}a_k}$ and
$\p^n\p_{a_1}\dots\p_{a_k}\G$. From this point of view, the last term is an
instanton, which vanishes if $\G$ is a linear function of $u$, that is, for
the $A_2$ and $\CP^1$-models.

One calculates that $\PHI_{2,1}$ is given by the explicit formula
\begin{align*}
\PHI_{2,1} &= \tfrac{1}{576} \bigl(
\half \, (\p u)^2 \, \G''' + \tfrac{9}{5} \, (\p u)^2 \, \G''
\, \G' + \tfrac{13}{5} \, \p^2u \, \G''
+ \tfrac{7}{10} \, \p^2u \, (\G')^2 \\
&- ( (\p v)^2 + \tfrac{7}{5} \, (\p u)^2 ) \, (\p u)^2 \, \phi' \,
\G'' \, \G' \, \Det^{-1} + \tfrac{6}{5} \, (\p u)^4 \, \phi'' \,
(\G')^2 \, \Det^{-1} \\
&+ ( \tfrac{2}{5} \, ( \p^2v \, \p v - \p^2u \, \p v ) \, \p v -
\tfrac{1}{10} \, (\p u)^4 \, \phi'' ) \, \G'' \, \Det^{-1} \\
&+ ( \tfrac{12}{5} \, \p^3v \, \p v - \tfrac{12}{5} \, \p^3u \, \p u \,
\phi'  - \tfrac{7}{5} \, \p^2u \, (\p u)^2 \, \phi'' ) \, \G' \,
\Det^{-1} \\
&+ \tfrac{11}{5} ( 4 \, \p^2v \, \p^2u \, \p v \, \p u \, \phi' - ( \p^2v
+ \p^2u \, \phi' ) ( (\p v)^2 + (\p u)^2 \, \phi' ) \\
&\quad + 2 \, ( \p^2v \, \p u - \p^2u \, \p v ) \, (\p u)^2 \, \p v \,
\phi'' - \half \, (\p u)^6 \, (\phi'')^2 ) \, \G' \, \Det^{-2} \bigr) .
\end{align*}

Now let $\PHI_2$ be a general solution of \eqref{TRR21} and
\eqref{TRR22}. Write $\PHI_2=\PHI_{2,0}+\PHI_{2,1}+f_2$. By the equations
\eqref{TRR21}, $D_{k,a}f_2=0$ for $k>1$; thus, $f_2$ is a function of the
coordinates $\{u,\p v,\p u\}$.
\begin{theorem} \label{H}
Define the functions $h_a=h_a(u)$ by the formula $h_a = \dfrac{\p f_2}
{\p(\p u^a)}\Big|_{(\p v,\p u)=(1,0)}$.  Then
$$
f_2 = \half \, \p u^a \, \p\UU_a^b \, h_b = \half \, \p u^a \, \p u^b \,
\AA_{ab}^c h_c
= \half \bigl( (\p v)^2 + \phi'\,(\p u)^2 \bigr) \, h_0(u) +
\p v\,\p u\,h_1(u) .
$$
\end{theorem}
\begin{proof}
Let $\tilde{f}_2=\half \, \p u^a \, \p\UU_a^b \, h_b$; then
$D_{1,1,a,b}\tilde{f}_2=0$ and $h_a = \dfrac{\p \tilde{f}_2}{\p(\p
u^a)}\Big|_{(\p v,\p u)=(1,0)}$.

Thus $f_2-\tilde{f}_2$ satisfies the equations
$\eth_{k,a}(f_2-\tilde{f}_2)=0$ for $k>1$, and
$\eth_{1,1,a,b}(f_2-\tilde{f}_2)=0$, as well as the dilaton equation
$\Dil(f_2-\tilde{f}_2)=2(f_2-\tilde{f}_2)$, and is thus determined by the
restrictions of the partial derivatives $\p_{1,a}(f_2-\tilde{f}_2)$ to the
small phase space. But these vanish; we conclude that $f_2=\tilde{f}_2$.
\end{proof}

In the next section, we determine the functions $h_a$.

\section{Virasoro constraints}

We now show that the Virasoro constraints $L_0Z=L_1Z=0$ of Eguchi, Hori and
Xiong \cite{EHX}, as generalized to arbitrary Frobenius manifolds by
Dubrovin and Zhang \cite{DZ0}, may be used to complete the determination of
the genus $2$ potential in two-primary models of topological gravity.

\subsection*{The constraint $L_0Z=0$}
According to Dubrovin and Zhang \cite{DZ0}, an Euler vector on a Frobenius
manifold determines matrices $\mu$ and $R[n]$, $n>0$, which satisfy the
commutation relations $[\mu,R[n]]=n\,R[n]$ and the symmetry conditions
$\mu_{ab}+\mu_{ba}=0$ and $$
R[n]_{ab}+(-1)^nR[n]_{ba}=0 .
$$
The basis $\Tau_a$ of primary fields may be chosen in such a way that the
matrix $\mu$ is diagonal
$$
\mu_a^b=\delta_a^b\,\mu_a ,
$$
and $\mu_0<\mu_a$ for $a\ne0$. Setting $d_a=\mu_a-\mu_0$ and $d=-2\,\mu_0$,
we have $\mu_a=d_a-d/2$.

For the Gromov-Witten invariants of a K\"ahler manifold $X$, the primaries
$\Tau_a$ form a basis of the De Rham cohomology $H^*(X,\C)$, and the number
$d_a$ is the holomorphic degree of $\Tau_a$, that is $\Tau_a\in
H^{d_a,*}(X,\C)$. (In particular, $d$ equals the complex dimension of $X$.)
In this case, $R[1]$ is the matrix of multiplication by $c_1(X)$, and
$R[n]=0$ for $n>1$.

Introduce the vector field
$$
\CL_0 = \sum_{k=0}^\infty \Bigl\{ ( \mu_a^b+k+\half ) \tt_k^a \, \p_{k,b} +
\sum_{\ell=1}^k R[\ell]_a^b \, \tt_k^a \, \p_{k-\ell,b} \Bigr\} ,
$$
where $\tt_k^a$ are the shifted coordinates
$\tt_k^a=t_k^a-\delta_{k,1}\delta_0^a$. The Virasoro constraint $L_0Z=0$ in
genus $g=0$ may be expressed as the following equation:
\begin{equation} \label{L00}
\CL_0\UU + \UU + [\mu,\UU] + R[1] = 0 .
\end{equation}
In genus $g>0$, the Virasoro constraint $L_0Z=0$ says that
\begin{equation} \label{Hori}
\CL_0 \PHI_g + \tfrac{1}{4} \, \delta_{g,1} \Tr(\tfrac14-\mu^2) = 0 .
\end{equation}
These equations are known to hold for Gromov-Witten invariants \cite{Hori}.

Let $\Euler=\Euler^a\p/\p u^a$ be the Euler vector field, where
\begin{equation} \label{Euler}
\Euler^a = (1-d_a) u^a + R[1]_0^a .
\end{equation}
Then \eqref{L00} implies that
\begin{equation} \label{L000}
\CL_0 u^a + \Euler^a = 0 .
\end{equation}
In calculating the action of the vector field $\CL_0$ in the coordinate
system $\{u^a_n\}$, we use \eqref{L000} together with the commutation
relation $[\p,\CL_0]=\half(1-d)\p$.

In the case of two primary fields, we have $\mu = \frac{1}{2} \bigl[
\begin{smallmatrix} - d & 0 \\ 0 & d \end{smallmatrix}\bigr]$. Consider
first the case in which $d$ equals $1$; then $R[1] = \bigl[
\begin{smallmatrix} 0 & r \\ 0 & 0 \end{smallmatrix} \bigr]$. By
\eqref{L00}, we see that $\phi=c\,e^{2u/r}$; redefining $u$, we may assume
that $r=2$, and we recover the $\CP^1$-model. Since
$\Tr(\mu^2-\tfrac14)=0$, we see from \eqref{Hori} that $\CL_0\PHI_1=0$;
\eqref{elliptic}, now shows that $\psi=-\tfrac{1}{24}u$, consistent with
the known form of $\PHI_1$ in the $\CP^1$-model.

The equation $\CL_0\PHI_2=0$ of \eqref{Hori} constrains the functions
$h_a(u)$ of Theorem \ref{H}; if $d=1$, it forces them to have negative
degree in $e^u$, and hence to vanish, as we have already observed,

If $d\ne1$, the matrix $R[1]$ vanishes. By \eqref{L00}, we see that
$\phi(u)=u^{(1+d)/(1-d)}$, up to a constant which we take to equal
$1$. (For example, the $A_2$-model, has $d=\tfrac13$ and $\phi(u)=u^2$.)
In genus $1$, the equation \eqref{Hori} shows that $\psi(u)$ is
proportional to $\log(u)$; both \eqref{elliptic} and \eqref{Hori} yield the
same answer for this constant,
$$
\psi(u) = \frac{d(3d-1)}{24(d-1)} \, \log(u) .
$$
Note that the $A_2$-model, for which $d=\frac13$, has $\psi=0$. The
equation $\CL_0\PHI_2=0$ imposes the homogeneities
$h_a(u)=C_a\,u^{((1+a)d-3)/(1-d)}$.

\subsection*{The constraint $L_1Z=0$}
Let $\CL_1$ be the vector field
\begin{align*}
\CL_1 &= - (\mu_a-\half)(\mu_a+\half) \<\<\Tau^a\>\>_0
\p_{0,a} + \sum_{k=0}^\infty \Bigl\{ ( \mu_a+k+\half ) ( \mu_a+k+\tfrac32 )
\tt_k^a \p_{k+1,a} \\
&+ \sum_{\ell\le k+1} 2(\mu_a+k+1) R[\ell]_a^b \tt_k^a \p_{k+1-\ell,b}
+ \sum_{\ell_1+\ell_2\le k+1} (R[\ell_1]R[\ell_2])_a^b \tt_k^a
\p_{k+1-\ell_1-\ell_2,b} \Bigr\} .
\end{align*}
Let $\VV=\Euler\UU$; by \eqref{L00}, $\VV=\UU+[\mu,\UU]+R[1]$. The
constraint $L_1Z=0$ in genus $0$ may written (Dubrovin and Zhang
\cite{DZ0}; cf. Theorem~5.7 of \cite{virasoro})
\begin{equation} \label{L10}
\CL_1 \UU + \VV^2 = 0 .
\end{equation}
In particular, we see that
\begin{equation} \label{L100}
\CL_1 u^a + \Euler^b \Euler^c \AA_{bc}^a = 0 .
\end{equation}
In genus $g>0$,the constraint $L_1Z=0$ is
\begin{equation} \label{L1}
\CL_1\PHI_g + \half \, ( \tfrac14 - \mu^2 )^{ab} \biggl( \sum_{h=1}^{g-1}
\<\<\Tau_a\>\>_h \, \<\<\Tau_b\>\>_{g-h} + \<\<\Tau_a\Tau_b\>\>_{g-1}
\biggr) = 0 .
\end{equation}
In the case of two primaries, this becomes
\begin{equation} \label{L12p}
\CL_1\PHI_g + \tfrac18 \, ( 1 - d^2 ) \biggl( \sum_{h=1}^{g-1}
\<\<\Tau_0\>\>_h \, \<\<\Tau_1\>\>_{g-h} + \<\<\Tau_0\Tau_1\>\>_{g-1}
\biggr) = 0 .
\end{equation}

In calculating the action of the vector field $\CL_1$ in the coordinate
system $\{u^a_n\}$, we use \eqref{L100} and the commutation relation
\begin{equation} \label{L1p}
[\p_{0,a},\CL_1] = \bigl( (\mu+\half)(\mu+\tfrac{3}{2}) \bigr){}_a^b
D_{1,b} + \bigl( (\mu+\half)\VV+\VV(\mu+\half) \bigr){}_a^b D_{0,b} .
\end{equation}
In the case of two primaries, this implies that
$$
[\p,\CL_1] = \begin{cases} (1-d) \bigl( \tfrac{1}{4} \, (3-d) \, D_{1,0} +
v \, D_{0,0} + u \, D_{0,1} \bigr) , & d\ne1 , \\ 2 \, D_{0,1} , & d=1 .
\end{cases}$$

Using these formulas, we see that the case $g=2$ of \eqref{L12p} yields the
equation
\begin{multline*}
0 = \CL_1\PHI_2 + \tfrac14 (1-d^2) \bigl( \<\<\Tau_0\>\>_1 \,
\<\<\Tau_1\>\>_1 + \<\<\Tau_0\Tau_1\>\>_1 \bigr) \\
\begin{aligned}
= & - 6 \bigl( (d+1)C_0 + \tfrac{1}{5760} \, d(3d-1)(3d-5)(d-2) \bigr) \,
u^{-2} \, \p v \, \p u \\
& + 3 \, C_1 (d-1) u^{(d-2)/(1-d)} ( (\p v)^2 + \phi' \, (\p u)^2 ) .
\end{aligned}
\end{multline*}
It follows that $h_1 = 0$ and
\begin{equation} \label{h}
h_0 = - \frac{d(3d-1)(3d-5)(d-2)}{5760(d+1)} \, u^{(d-3)/(1-d)} .
\end{equation}
completing the determination of $\PHI_2$.

Our formula for $\PHI_2$ agrees with that of Dubrovin and Zhang \cite{DZ},
who apply the method of Eguchi and Xiong \cite{EX}; in other words, they
use the constraints $D_{k,a}\PHI_2=0$, $k>4$, and $L_nZ=0$, $n\le10$.

\subsection*{The higher Virasoro constraints}
The higher Virasoro constraints are given by formulas involving a Lie
algebra of vector fields $\CL_n$, $n\ge-1$, on the large phase space, which
satisfy the commutation relations
$$
[\CL_m,\CL_n] = (m-n) \CL_{m+n} .
$$
This Lie algebra is generated by $\CL_{-1}$ and $\CL_n$, for any $n>1$.

Just as for $\CL_0$ and $\CL_1$, we can avoid using the explicit formula
for $\CL_n$. The Virasoro constraint $L_nZ=0$ in genus $0$ may be written
\begin{equation} \label{Ln0}
\CL_n \UU + \VV^{n+1} = 0 .
\end{equation}
In calculating the action of the vector field $\CL_2$ in the coordinate
system $\{u^a_n\}$, we use \eqref{Ln0} and the commutation relation
\cite{Virasoro}
\begin{equation} \label{Lnp}
[\p_{0,a},\CL_n] = \sum_{i=0}^n (\BB_{n,i})_a^b D_{i,b} ,
\end{equation}
where the matrices $\BB_{n,i}$ are determined by the recursion
$$
\BB_{n,i} = (\mu+i+\half)\BB_{n-1,i-1} + \VV\,\BB_{n-1,i} ,
$$
with initial condition $\BB_{-1,i}=\delta_{i+1,0}$.

In the case of two primaries, with $n=2$, this implies that
$$
[\p,\CL_2] = \begin{cases} (1-d) \bigl( \tfrac{1}{8} (3-d)(5-d) D_{2,0} +
\tfrac{3}{4} (3-d) v D_{1,0} + \tfrac{1}{4} (d^2-2d+9) u D_{1,1} & \\ + (
\tfrac{3}{2} v^2 + \half (1+d)(3-d) u \,\phi(u) ) D_{0,0} + 3 uv D_{0,1}
\bigr) , & d\ne1 , \\[5pt] 6 \, D_{1,1} + 4 e^u D_{0,0} + 6 v D_{0,1} , &
d=1 .
\end{cases}$$

In genus $g>0$,the constraint $L_2Z=0$ is
\begin{align*} \label{L2}
\CL_2\PHI_g & + (\mu_a-\tfrac{3}{2})(\mu_a-\half)(\mu_a+\half)
\eta^{ab} \biggl( \sum_{h=1}^{g-1} \<\<\tau_{1,a}\>\>_h \,
\<\<\Tau_b\>\>_{g-h} + \<\<\tau_{1,a}\Tau_b\>\>_{g-1} \biggr) \\
& - \half \, (3\mu_a^2+3\mu_a-\tfrac{1}{4}) R[1]^{ab}
\biggl( \sum_{h=1}^{g-1} \<\<\Tau_a\>\>_h \, \<\<\Tau_b\>\>_{g-h} +
\<\<\Tau_a\Tau_b\>\>_{g-1} \biggr) = 0 .
\end{align*}
It may be verified that $\PHI_2$ satisfies this equation.

Since the differentials operators $L_n$, $n>-1$, lie in the Lie algebra
generated by $L_{-1}$ and $L_2$, it follows that the Virasoro conjecture
holds to genus $2$ for two-primary models.

\subsection*{The Belorousski-Pandharipande equation}

The Belorousski-Pandharipande equation \cite{BP} is a differential equation
satisfied by the genus $2$ potential, analogous to the equation
\eqref{elliptic} in genus $1$; it may be expressed as saying that a certain
cubic polynomial in the coordinates $\{u^a\}$ vanishes. It turns out that
in the case of backgrounds with two primaries, the equation gives a second
(and thus rigorous) derivation of the above formula for $C_0$, but leaves
$C_1$ undetermined.

Taking Theorem \ref{H} and the equations $\CL_{-1}\PHI_2=0$ and
$\CL_0\PHI_2=0$ into account, the Belorousski-Pandharipande equation
reduces to a single equation
$$
\phi' \, h_0' - \half \, \phi'' \, h_0 - \tfrac{1}{48} \, \G'''' -
\tfrac{3}{5} \, \G''' \, \G' + \tfrac{9}{10} \, (\G'')^2 = 0 .
$$
With $\phi(u)=u^{(1+d)/(1-d)}$ and
$\psi(u)=\frac{d(3d-1)}{24(d-1)}\,\log(u)$, the function $h_0$ of \eqref{h}
satisfies this equation.

\section*{Acknowledgments}

The authors thank B. Dubrovin and Y. Zhang for communication of their
results \cite{DZ} prior to publication, and for a number of interesting
conversations. The second author thanks Y. Zhang and Qinghua University for
its hospitality while working on this paper.

The research of the first author is supported in part by special priority
area \#707 of the Japanese Ministry of Education. The research of the
second author is supported in part by NSF grant DMS-9704320, and by the
special year ``Geometry of String Theory'' of RIMS. The research of the
third author is supported in part by NSFC grant 19925521 and by a startup
grant from Beijing University.


\begin{thebibliography}{99}

\bibitem{BP} P. Belorousski and R. Pandharipande, \emph{A descendent
relation in genus 2.} \texttt{<math.AG/9803072>}.

\bibitem{DW} R. Dijkgraaf and E. Witten, \emph{Mean field theory,
topological field theory, and multi-matrix models,}
Nucl. Phys. \textbf{B342} (1990), 486--522.

\bibitem{DZ0} B. Dubrovin and Y. Zhang, \emph{Frobenius manifolds and
Virasoro constraints} Selecta Math. (N.S.) \textbf{5} (1999),
423--466. \texttt{<math/9808048>}.

\bibitem{DZ} B. Dubrovin and Y. Zhang, in preparation.

\bibitem{EY} T. Eguchi and S.-K. Yang, \emph{The topological $\CP^1$ model
and the large-N matrix integral,} Mod. Phys. Lett. \textbf{A9} (1994)
2893--2902. \texttt{<hep-th/9407134>}

\bibitem{EHX} T. Eguchi, K. Hori and C.-S. Xiong, \emph{Quantum cohomology
and Virasoro algebra,} Phys. Lett. \textbf{B 402} (1997), 71--80.
\texttt{<hep-th/9703086>}

\bibitem{EHY} T. Eguchi, K. Hori and S.-K. Yang, \emph{Topological
$\sigma$-models and large-$N$ matrix integral,} Internat. J. Modern Phys. A
\textbf{10} (1995), 4203--4224. \texttt{<hep-th/9503017>}

\bibitem{EYY} T. Eguchi, Y. Yamada and S.-K. Yang, \emph{On the genus
expansion in the topological string theory,} Rev. Math. Phys. \textbf{7}
(1995), 279--309. \texttt{<hep-th/9405106>}

\bibitem{EX} T. Eguchi and C.-S. Xiong, \emph{Quantum cohomology at higher
genus: topological recursion relations and Virasoro conditions,}
Adv. Theor. Math. Phys. \textbf{2} (1998),
219--228. \texttt{<hep-th/9801010>}

\bibitem{elliptic} E. Getzler, \emph{Intersection theory on
$\overline{\mathcal{M}}_{1,4}$ and elliptic Gromov-Witten invariants},
J. Amer. Math. Soc. \textbf{10} (1997),
973--998. \texttt{<alg-geom/9612004>}

\bibitem{genus2} E. Getzler, \emph{Topological recursion relations in genus
$2$.} In ``Integrable systems and algebraic geometry (Kobe/Kyoto, 1997),''
World Sci. Publishing, River Edge, NJ, 1998,
pp. 73--106. \texttt{<math/9801003>}

\bibitem{virasoro} E. Getzler, \emph{The Virasoro conjecture for
Gromov-Witten invariants,} ``Algebraic geometry: Hirzebruch 70 (Warsaw,
1998),'' Contemp. Math. \textbf{241}, Amer. Math. Soc., Providence, RI,
1999, pp.  147--176.

\bibitem{Virasoro} E. Getzler, in preparation.

\bibitem{Hori} K. Hori, \emph{Constraints for topological strings in $d\ge
1$}, Nucl. Phys. \textbf{B439} (1995), 395--420. \texttt{<hep-th/9411135>}

\bibitem{Manin} Yu. Manin, ``Frobenius manifolds, quantum cohomology, and
moduli spaces,''
Amer. Math. Soc. Colloq. Publ. \text{47}. Amer. Math. Soc., Providence, RI,
1999.

\bibitem{P} R. Pandharipande, \emph{The Toda equations and the
Gromov-Witten theory of the Riemann sphere.} \texttt{<math/9912166>}

\bibitem{Witten} E. Witten, \emph{Two dimensional gravity and
intersection theory on moduli space}, Surveys in Differential
Geom. \textbf{1} (1991), 243--310.

\end{thebibliography}
\end{document}